\documentclass[epjST]{svjour}
\usepackage{graphics}
\usepackage{epsfig}
\begin{document}
\title{Role of coherence in resistance quantization}
\subtitle{Quantum Hall and quantum point contact versus charge relaxation resistance}
\author{Markus B{\"u}ttiker\inst{1}\fnmsep\thanks{\email{Markus.Buttiker@physics.unige.ch}}
\and Simon E. Nigg\inst{1}
}
\institute{D{\'e}partement de Physique Th{\'e}orique, Universit{\'e} de Gen{\`e}ve, CH-1211 Gen{\`e}ve, Switzerland}
\abstract{
The quantization of resistances in the quantum Hall effect and ballistic transport through quantum point contacts is compared with the quantization of the charge relaxation resistance of a coherent mesoscopic capacitor. While the former two require the existence of a perfectly transmitting channel, the charge relaxation resistance remains quantized for arbitrary backscattering. The quantum Hall effect and the quantum point contact require only local phase coherence. In contrast quantization of the charge relaxation resistance requires global phase coherence. 
} 
\maketitle
\section{Introduction}
\label{intro}

\begin{figure}
\resizebox{0.75\columnwidth}{!}{%
\centerline{\includegraphics[width=10cm]{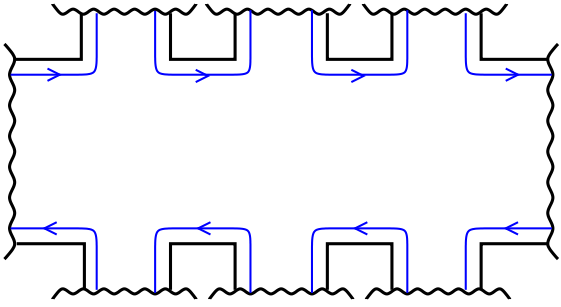}} }
\caption{Hall bar with three pairs of voltage probes. Edge states following sample boundaries are indicated schematically with their chirality marked by arrows. Well connected voltage probes completely suppress phase coherence. 
}
\label{fig:1}      
\end{figure}

Several electrical transport phenomena are known which at sufficiently low temperatures exhibit quantized resistances. The most famous of these phenomena is the quantum Hall effect which can be observed in a Hall bar patterned in a two dimensional electron gas (see Fig. \ref{fig:1}). In the presence of a magnetic field applied perpendicular to the plane of the two-dimensional electron gas, the Hall resistance is found to exhibit plateaus at values \cite{intqhe} 
\begin{equation}
\label{eq1}
R_{H} = \frac{h}{e^2} \frac{1}{N}, 
\end{equation} 
where $h$ is Planck's constant and $e$ is the elementary charge. The precision can be so high that this effect is now used to determine the resistance standard \cite{jeanneret}.
Simultaneously with the quantization of the Hall resistance the longitudinal resistance vanishes.
Subsequent to the discovery of the integer quntum Hall effect many additional plateaus where found at fractional values \cite{fracqhe}. Even within the integer quantum Hall effect, patterning the two-dimensional electron gas with the help of gates such that transport is through regions with different filling factors, leads to quantized longitudinal resistances and Hall resistances at fractions determined by the filling factors of the different regions \cite{buttiker88,washburn88,haug88}.

\begin{figure}
\resizebox{0.75\columnwidth}{!}{%
\centerline{\includegraphics[width=10cm]{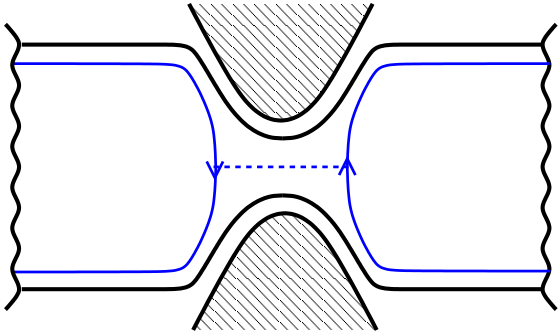}} }
\caption{Quantum point contact connecting two two-dimensional electron gases with a metallic contact to the left and right. The shaded areas mark a top gate which permits to tune the opening of the contact. A single edge state is shown for the case of partial transmission. 
}
\label{fig:2}      
\end{figure}

Interestingly, ballistic transport through a quantum point contact \cite{ball1,ball2}, a small opening connecting two large banks of two-dimensional electron gas (see Fig. \ref{fig:2}), exhibits resistance plateaus at 
\begin{equation}
\label{eq2}
R = \frac{h}{2e^2} \frac{1}{N}.
\end{equation}
Here the magnetic field is zero and a factor two appears since spin degeneracy is not lifted. Within scattering theory the explanation of the quantization of resistances in these two transport problems is the same.

It is the purpose of this article to compare these quantization phenomena with a resistance quantization which has been predicted \cite{btp93} and observed \cite{gabelli06} in the dynamic response of a capacitor. A capacitor does not support a dc-current but when charged exhibits a relaxation current with a relaxation time determined by its capacitance and the resistance in series. For a mesoscopic capacitor the capacitance $C_{\mu}$ is a sample specific quantity but, in the limit that the connection to the mesoscopic capacitor plate is through a single spin polarized channel (see Fig. \ref{fig:3}), its resistance $R_{q}$ is independent of the properties of this connection and given 
by the universal value \cite{btp93}
\begin{equation}
\label{eq3}
R_q = \frac{h}{2e^2}.
\end{equation} 
Note that Eq. (\ref{eq3}) and Eq. (\ref{eq2}) for $N = 1$ are the same. However, whereas for the quantum point contact the factor $2$ arises from spin degeneracy, Eq. (\ref{eq3}) is for a spin polarized channel and the factor $2$ has an entirely different origin. Comparison of these transport phenomena might lead to a better understanding of the necessary conditions for resistance quantization.

\section{Scattering theory of quantized resistance}
\label{scattering}

\begin{figure}
\resizebox{0.75\columnwidth}{!}{%
\centerline{\includegraphics[width=10cm]{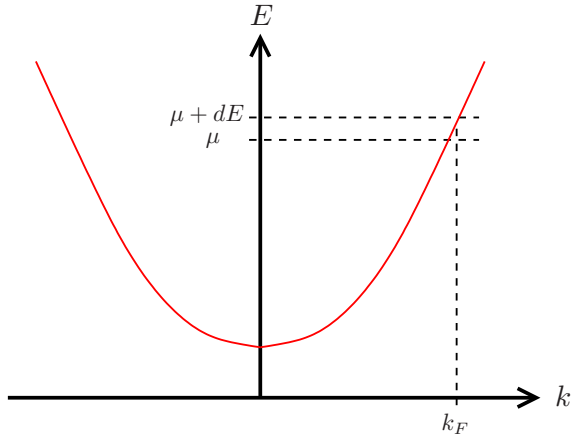} }}
\caption{Dispersion relation $E(k)$ of a (magneto-elastic) subband. The current of right movers in the energy interval $\mu$ and $\mu + dE$ is independent of the properties of the channel and given by 
$(e/h) dE$ with $e$ the electron charge and $h$ Planck's constant. 
}
\label{fig:4}      
\end{figure}

Discussions which view the sample as a scatterer that permits transmission and reflection \cite{landauer,anderson,imry} of carriers lead to quantized conductance (resistance) due to two basic properties. First scattering theory formulates a conduction problem in terms of scattering channels. In an asymptotic region (the lead, the reservoir) the Schroedinger equation is assumed to be separable into longitudinal (along $x$) and transverse motion (along $y$) with wave function $\exp(ikx) \chi_{k}(y)$. Such a state has an energy $E(k)$ composed of the kinetic energy for motion along $x$ and a confinement energy (see Fig. \ref{fig:4}). Of relevance are the properties of this quantum channel at the electrochemical potential $\mu$ (the Fermi energy plus the electrostatic potential).  In particular we are interested in the current that can be carried by such a channel in a small energy interval $\mu, \mu + dE$ as shown in Fig. \ref{fig:4}. The current is $dI = e v  d\rho$, where 
$v$ is the carrier velocity at energy $\mu$ and $d\rho$ is the density of carriers with energy in the interval. The velocity is $v = \hbar^{-1} dE/dk$ and the density is $d\rho = (d\rho/dk) (dk/dE) dE$.  Since for a one-dimensional channel $(d\rho/dk) = 1/2\pi$ and since $(dk/dE) = (1/\hbar v)$ we arrive at the important result 
\begin{equation}
\label{eq4}
dI = \frac{e}{h} dE\,.
\end{equation} 
The current of left movers in a quantum channel depends only on fundamental constants $e$ and $h$ and on the width of the energy interval. The current is independent of the properties of the channel. Eq. (\ref{eq4}) assumes that all states are fully occupied. Thus Eq. (\ref{eq4}) gives the maximum current
that can be carried by states in this energy interval.

\begin{figure}
\resizebox{0.75\columnwidth}{!}{%
\centerline{\includegraphics[width=10cm]{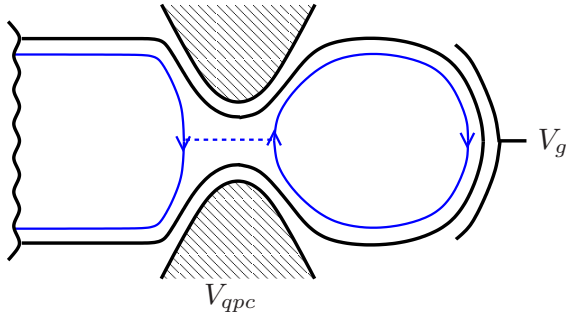} }
}
\caption{Mesoscopic capacitor. A cavity is connected via a quantum point contact to a two-dimensional electron gas which couples to a metallic contact. The cavity is coupled capacitively to a gate at voltage $V_g$. The opening of the quantum point contact can be tuned with the voltage $V_{QPC}$.  
}
\label{fig:3}      
\end{figure}

In scattering theory Eq. (\ref{eq4}) determines the current that is incident on the sample in the zero temperature limit if a voltage $\mu + dE - \mu = eV$ is applied to the sample. We then immediately have a quantized conductance if, and this is the second important condition that must be fulfilled, 
no forward moving carriers are scattered into the backward direction of the same channel or into any other channel that might be present. This is the condition of no backscattering \cite{buttiker88}. Equivalently we might say that scattering channels are chiral, they permit motion only into one direction. 

We emphasize that in a realistic description fully transmitted channels and fully reflected channels co-exist. This is clearly the case in a quantum point contact, where we have a large number of channels in the wide region but only very few channels in the center of the constriction. It is also the case in the quantum Hall effect, where we have a large number of channels in a metallic contact but only a few channels in the semiconducting part of the structure. In fact, typically, the property that each channel is either fully reflected or fully transmitted when conductance is quantized will be true only in a special basis (the basis of eigenchannels).

These conditions are equally valid for a quantum point contact at zero magnetic field. However at high magnetic fields the quantum channels which we have introduced above become edge states \cite{halperin82} corresponding to quantized orbits describing skipping of carriers along the sample boundary. The edge channels are indicated in Figs. \ref{fig:1},\ref{fig:2},\ref{fig:3} and their chirality is marked by arrows. Therefore, the channels can now be visualized in real space. For small applied voltages the edge channels provide unidirectional transport along sample boundaries from one sample contact to another. Backscattering would imply a hopping from an edge state to localized states (not shown) in the center of the bar all the way across to the other edge of the sample. In a high magnetic field tunneling over distances large compared to the magnetic length $l_B = (hc/|eB|)^{1/2}$ is exponentially suppressed. At low temperature, the exponential suppression of backscattering accounts for the high accuracy of the resistance quantization of the quantum Hall effect \cite{buttiker88}. 

There is still occasionally some skepticism expressed concerning the significance of edge states. 
On the other hand in a portion of the literature edge states are taken to be self-evident. As a matter of fact, it was only the notion that non-ideal contacts lead to a non-equilibrium population of edge states \cite{buttiker88,komiyama89,vanwees89,alphenaar90} and the detection of these different populations which made edge states a physical reality.

How quantum are quantized resistances? To be more precise, over what distances must electron motion be phase coherent? It is clear that the quantized Hall effect can be measured in samples that are much larger than an electron phase-coherence length. 
To derive Eq. (\ref{eq4}) we only needed two inputs: The velocity of the channel and the fact that the density of states is inversely proportional to the velocity. The velocity is determined by the dispersion relation of the channel. This dispersion relation contains essentially only short range properties of the sample. A mechanism which would destroy the phase memory of a carrier does not affect the carrier velocity. Dephasing will broaden the singularity of the density of states near the channel threshold but have little effect for carrier energies above threshold. Clearly, global phase coherence is not needed to arrive at Eq. (\ref{eq4}). Of course, for the quantized Hall effect we need a Landau level structure, respectively, edge channels with a chiral property. However, to establish locally a Landau level structure it is sufficient to have a phase coherence length of a few cyclotron radii. 

Indeed, it is well known that we can place an arbitrary number of voltage probes along the perimeter of a Hall bar without generating deviations from quantization. Each voltage probe absorbs carriers and reinjects them with a phase that is unrelated to the phase of absorbed carriers. Voltage probes that fully connect to an edge channel completely dephase electron motion along the sample boundary. Nevertheless we have perfect quantization of the Hall resistance. In fact, it has been pointed out, that inelastic scattering (a process which breaks not only the phase but also relaxes carriers in energy), can actually help to quantize Hall resistances in cases where in the absence of such relaxation processes the Hall resistance would exhibit deviations from the quantized value  \cite{buttiker88}. In particular, voltage probes can provide inelastic relaxation. It is very likely for this reason that manuals for a high precision measurement \cite{jeanneret} call for a Hall bar with three pairs of voltage probes, and ask for the Hall voltage to be measured at the middle pair. 

Similarly for quantum point contacts: for subbands to be defined, it is sufficient that electron motion inside the contact is phase coherent over a time it takes a carrier to travel back and forth between the confining walls of the conductor. Again, for quantization of conductance in a quantum point contact, long range phase coherence is not necessary. 

Interestingly, as we will discuss below, the requirements for the quantization of the charge relaxation resistance in a mesoscopic capacitor are entirely different. Arbitrary backscattering is possible but global phase coherence is necessary.
 
\section{Quantized charge relaxation resistance}
\label{rq}

Fig. \ref{fig:3} shows a schematic picture of a mesoscopic capacitor \cite{btp93,ptb96}. A small cavity, forming one plate of the capacitor, is via a quantum point contact connected to a metallic reservoir \cite{gabelli06}. The opening of the quantum point contact can be adjusted with the gate voltage $V_{QPC}$.
A side gate at voltage $V_g$ forms the other plate of the capacitor. We assume that a strong magnetic field is applied perpendicular to the plane of the two-dimensional electron gas. Carriers incident form the metallic contact follow an edge state and at the entrance to the cavity are reflected with amplitude $ir$ (probability $R = |r|^2$ ) and can transmit through the quantum point contact with amplitude $t$ (transmission probability $T = |t|^2$ ). Inside the cavity carriers follow a circular edge state. During each cycle they gain a phase $\phi$.
Upon reaching the quantum point contact they can leave the cavity with amplitude $t$ or continue for an additional cycle with reflection amplitude $ir$. The overall scattering matrix $s$ is a Fabry-Perot type expression which sums up the amplitudes of all the possible scattering events, 

\begin{equation}
\label{eq5}
s (\epsilon) = e^{i\phi} \frac{1 + i r e^{-i\phi}}{1 - i r e^{i\phi}} \,.
\end{equation}
The absolute value is $|s| = 1$. Each incident carrier is eventually reflected with probability 1. The capacitor does not support a dc-current but an ac-current can be induced by applying an oscillating voltage between the metallic contact and the side gate. 

Classically the structure resembles a circuit consisting of a capacitance $C$ and a  resistance $R$ (the quantum point contact) in series. The admittance $G(\omega) = dI(\omega)/dV(\omega)$ has a low frequency expansion 
\begin{equation}
\label{eq6}
G(\omega) = -i \omega C + \omega^{2} C^{2} R + ....\,\,
\end{equation}
For a mesoscopic capacitor a similar expansion holds 
\begin{equation}
\label{eq7}
G(\omega) = -i \omega C_{\mu} + \omega^{2} C_{\mu}^{2} R_{q} + ....\,\,
\end{equation}
but now the capacitance $C_{\mu}$ and charge relaxation resistance $R_{q}$ are very different from those of a classical circuit. 

Consider for a moment a conductor with a geometry similar to the one shown in Fig. \ref{fig:3} but allowing for many channels to enter and leave the cavity. Instead of a single channel, we now have a many channel scattering problem. The amplitudes of the reflected currents in the $M$ channels are determined by a scattering matrix $s$ with dimension $M \times M$. Let us further assume that we can, as is done in the theory of Coulomb blockade, describe the potential in the cavity by a single parameter $U$. The Coulomb energy is determined by a geometrical capacitance $C$. Below we give the results which apply when the interaction is treated in Hartree approximation. The central point we are making also holds if interactions are treated in Hartree-Fock \cite{nigg06,nano07}. For simplicity, we also consider throughout only the zero temperature limit. The scattering properties of the capacitor \cite{btp93} enter via the Wigner-Smith matrix $N = (2\pi i)^{-1} s^{\dagger} ds/dE$. The capacitance is the series addition of the geometrical capacitance and a quantum capacitance $C_{q} = e^{2} Tr[N]$,  
\begin{equation}
\label{eq8}
C^{-1}_{\mu} = C^{-1} +( e^{2} Tr[{N}])^{-1} .
\end{equation} 
The trace is over scattering channels. 
For weak interaction, large geometrical capacitance $C$, the capacitance $C_{\mu}$ is determined by the quantum capacitance $C_{q}$. For small capacitance the Coulomb charging energy is prohibitive and 
$C_{\mu} \approx C$ is very small. Mesoscopic fluctuations of the capacitance due to fluctuations of the density of states (the quantum capacitance) are the subject of Refs. \cite{gopar,brouwer}. 

The charge relaxation resistance is \cite{btp93} 
\begin{equation}
\label{eq9}
R_q  = \frac{h}{2e^{2}} \frac{Tr[N^{\dagger}N] }{(Tr[N])^{2}} = 
\frac{h}{2e^{2}} \frac{\sum_{n= 1}^{n= M}( d\phi_{n}/dE)^{2} }{(\sum_{n=1}^{n =M}d\phi_{n}/dE)^{2}}\,.
\end{equation} 
The second equality has been obtained by noting that the scattering matrix can be diagonalized and since we deal with reflection only the eigenvalues are of the form $e^{i\phi_{n}}$.

Eq. (\ref{eq9}) is valid for an $M$-channel scatterer. If we now go back to a situation where only one channel is (partially) transmitted at the quantum point contact $M =1$ we see immediately that 
Eq. (\ref{eq9}) leads to a quantized resistance given by Eq. (\ref{eq3}). Interestingly this resistance is independent of the degree of transmission through the contact. Unlike the quantized resistance in the integer quantum Hall effect or the quantized resistance of a quantum point contact it does not depend on the suppression of backscattering. 

However, whereas in the integer quantum Hall effect or in quantum point contacts we can observe a whole sequence of plateaus, here, for two or more channels, $R_q$ is typically not quantized. An exception occurs only if the channels are degenerate. For $K$ degenerate channels the resistance decreases with 
$K$ as $R_q =  (h/2e^{2})(1/K)$. For instance a single walled Carbon nanotube connecting two metallic contacts and coupled perfectly to them has $K = 8$ degenerate channels and the charge relaxation resistance is $R_q =  (h/16e^{2})$. Here we assumed that interactions do not split this degeneracy. 

Still another interpretation of the quantization of the charge relaxation resistance is offered if one 
multiplies the Wigner-Smith matrix by Planck's constant $h$ and regards its elements as time-delays, 
$\tau = -i \hbar s^{\dagger} ds/dE$. The charge relaxation resistance is then the mean square time-delay divided by the square of the time-delay, $R_q = (h/2e^{2})<\tau^2>/(<\tau >)^2$. Therefore, the quantization of the charge relaxation resistance is a consequence of the fact that for a single channel scattering problem, the mean square dwell time is equal to the square of the dwell time \cite{leavens}. 

For further discussion of charge relaxation resistances we mention work on gated ballistic wires \cite{blanter}, 
Carbon nanotubes connected to a single contact \cite{ww08} and in two terminal geometry \cite{burke} 
and the work on charge relaxation in fractional edge channels \cite{pham}. A recent discussion \cite{wang} of linear response 
also considers the term third order in frequency. Charge relaxation in the presence of large amplitude pertrubations has been examined experimentally \cite{feve} and theoretically \cite{moskalets}. 

\section{Decoherence and charge relaxation}
\label{rqdeph}

There is still another important difference between the quantization of $R_q$ and the quantization of the resistances in a ballistic quantum point contact or the integer quantum Hall effect. Quantum coherence is absolutely crucial, and it is crucial not only within the quantum point contact but even within the cavity. In a recent work \cite{nigg08} we have investigated the charge relaxation resistance when coherence within the cavity is destroyed. Dephasing is introduced with the help of an additional voltage probe which is connected to the cavity. A carrier leaving the cavity into the voltage probe is replaced by a carrier which enters the cavity from the voltage probe. Leaving and entering carriers have no phase relationship. A tunneling contact with transmission probability $\epsilon$ connects cavity and voltage probe. The transmission probability and dephasing time $\tau_{\phi}$ are related via $\epsilon = 1 - \exp(-h/\Delta \tau_{\phi})$. Here $\Delta$ is the level spacing in the cavity (if the quantum point contact is closed off).

\begin{figure}
\resizebox{0.75\columnwidth}{!}{%
\centerline{\includegraphics[width=10cm]{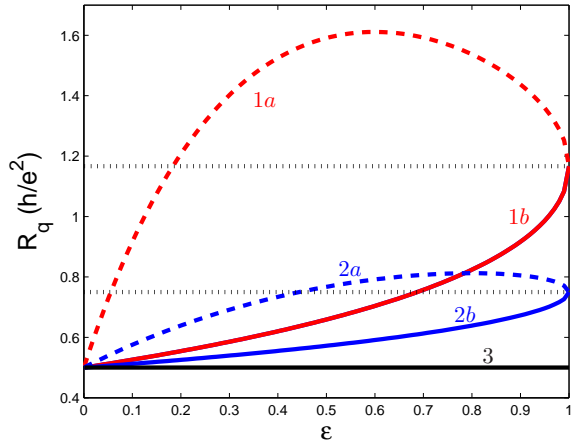}} 
}
\caption{Charge relaxation resistance as a function of decoherence strength $\epsilon$ for different transmission probabilities (1) $T = 0.6$, (2) $T = 0.8$ and (3) $T=1$ of the quantum point contact. For $\epsilon = 0$ the cavity is fully coherent, for $\epsilon =1 $ a carrier looses phase memory before completing a turn along the edge state. For a quantum point contact with $T=1$ there is no interference and $R_q$ remains quantized (curve 3).  If $T$ is smaller than $1$, $R_q$ evolves differently, depending on whether the Fermi energy is on resonance with a state in the cavity 
(full lines) or whether the Fermi energy is between two states in the cavity (dashed lines). After Ref. \protect\cite{nigg08}. 
}
\label{fig:5}      
\end{figure}

The limiting value (for $\epsilon = 1$) of the charge relaxation resistance is \cite{nigg08}
\begin{equation}
\label{eq10}
\lim_{\epsilon\rightarrow 1} R_q = \frac{h}{e^2} \left(\frac{1}{2} + \frac{1-T}{T}\right)\,.
\end{equation} 
Interestingly, for a completely dephased edge channel the charge relaxation resistance is the sum of 
a resistance ${h}/{2e^2}$ plus the (original) Landauer resistance $(h/{e^2})(1-T)/T$. Naively one might have expected that the charge relaxation resistance tends towards the two-terminal resistance 
$(h/{e^2})(1/T)$. That this is not the case indicates that full relaxation towards an equilibrium distribution is not achieved if there is a single incoherent state in the cavity. Indeed it can be shown\cite{nigg08} that if the model is extended to take into account localized states in the cavity and if inelastic scattering is permitted between the dephased edge state and such localized states, the cavity eventually starts to behave like a reservoir and the charge relaxation resistance does indeed tend toward $(h/{e^2})(1/T)$.

The central message is that here we have an example of a quantized resistance which does not require suppression of back scattering but which is very sensitive to dephasing. Quantization of the charge relaxation resistance needs coherence on a global scale. It is not sufficient to have phase coherence inside the quantum point contact but phase coherence over the quantum point contact plus the entire cavity is needed. 

\section{Interface resistances}
\label{rc}

The charge relaxation resistance $R_q = \frac{h}{2e^2}$ has the same value as the reservoir-lead interface resistance of a single channel wire. The interface resistance $R_c$ appears if we compare the Landauer resistance $R_L = (h/{e^2 })(1-T)/T$ and the two-probe resistance $R = (h/{e^2})(1/T)$. The difference is $(h/e^2 )$. This can be viewed \cite{imry,landauer} as the sum of two contact resistances $R_c = h/2e^2$ one for each reservoir-lead interface. Despite the fact that the two resistances are the same, the physics which leads to them is not. The contact resistance $R_c$, like the charge relaxation resistance $R_q$ are independent of the transmission probability of the scatterer. But while $R_q$ applies for a fully phase coherent system, the Landauer resistance $R_L$ neglects the effects of voltage probes needed to measure it. It is a result valid only in the limiting case that voltage probes couple weakly and phase coherence between scatterer and probes is destroyed \cite{mb89}. Therefore $R_c$ is really a result that applies in the absence of global phase coherence. The distinction between $R_q$ and $R_c$ becomes even more apparent when multichannel geometries are considered \cite{nigg08}. While $R_c$ depends only on density of states 
(derivatives of scattering phases with respect to potential or energy) the charge relaxation resistance contains terms which are quadratic in the density of states \cite{nigg08}. 

\section{Discussion and conclusion}
\label{dis}
Comparison of the quantized resistances of the Hall effect and the quantum point contact with the quantized charge relaxation resistance of a mesoscopic capacitor reveals two entirely different origins. Whereas the quantization of the Hall resistance and the resistance of quantum point contact are due to the existence of scattering channels with perfect transmission from one sample contact to another, quantization of the charge relaxation resistance rests on the fact that for a single scattering channel the mean squared dwell time is equal to the square of the dwell time. This latter quantization applies to one channel only (or to a set of degenerate channels). In contrast as long as transmission is perfect quantization can be observed for transport through an arbitrary number of channels. On a plateau the dc-response of a conductor in the quantized Hall regime is truly universal, at least for the four-probe resistances. In contrast, the response of a mesoscopic capacitor is sample specific, even when the charge relaxation resistance is quantized.

\section*{Acknowledgments}
\label{acknowledgment}
This work was supported by the Swiss National Science Foundation, the STREP project SUBTLE, and the Swiss National Center of Competence in Research, MaNEP.

\section*{Appendix}
\label{appendix}

We present here the expression for the charge relaxation resistance in terms of a self-consistent potential landscape. For a geometrical capacitance, in random phase approximation, $R_q$ does not depend explicitly on the interaction. However, in the more general case treated here, an explicit dependence on interaction is found. The results can be expressed in terms of an effecive Wigner-Smith (density of states) matrix. Consider a cavity with $M$ incident channels. The Wigner-Smith matrix is 
$N = (2\pi i)^{-1} s^{\dagger} ds/dE$. We introduce a local density of states matrix in which the energy derivative is replaced by a functional derivative with respect to the electrostatic potential $U(r)$, $n(r) = - (2\pi i)^{-1} s^{\dagger} \delta s/\delta eU(r)$. Its trace is the local density of states $\nu(r) = Tr[n(r)] = - (2\pi i)^{-1} Tr[s^{\dagger} \delta s/\delta eU(r)]$. Coulomb interaction leads to a screened charge 
described by an effective density of states and an effecitve density of states matrix \cite{mb96}
\begin{equation}
\label{eq11}
n^{eff}(r) = n(r) - \int d^3 r^{\prime}\, \nu(r)\, g(r,r^{\prime})\, n(r^{\prime})
\end{equation} 
Here $g(r,r^{\prime})$ is a an effective interaction. It is the Green's function of the Laplace 
equation with a non-local screening kernel $\pi(r,r^{\prime})$, 
\begin{equation}
\label{eq12}
\nabla^{2}\, u(r) +  4 \pi \, e^2 \, \int d^3 r^{\prime}\, \pi(r,r^{\prime})\, u(r^{\prime}) = 4 e^{2} \pi \,n(r), 
\end{equation} 
and determines the solution of the potential (matrix) $u(r) = \int d^3 r^{\prime} g(r,r^{\prime}) n(r^{\prime})$.
In terms of the effective density of states matrix the capacitance of the structure is 
\begin{equation}
\label{eq13}
C_{\mu} = e^{2}\, \int d^3 r\, Tr[n^{eff}(r)].
\end{equation} 
The charge relaxation resistance is 
\begin{equation}
\label{eq14}
R_q = \frac{h}{2e^{2}}\, \frac{ Tr[(\int d^3 r\, n^{eff}(r))^{\dagger}\,(\int d^3 r\, n^{eff}(r))]}{(\int d^3 r\, Tr[n^{eff}(r)])^{2}}.
\end{equation} 
To derive Eq. (\ref{eq14}) it is probably easiest to calculate the current noise and to use the fluctuation-dissipation theorem. We note that for our problem with reflection only, the current-operator in second quantization is, in the absence of interaction, to first order in $\omega$, 
\begin{equation}
\label{eq15}
\hat I (\omega) = -i\, e\, \omega\, \int dE\, \hat a^{\dagger}(E)\, n(r)\, \hat a(E). 
\end{equation} 
In the presence of interaction, we can again replace $n(r)$ by $n^{eff}$. Using the quantum statistical expectation values of the annihilation and creation operators for a free electron gas, 
and writing the noise spectrum in the form $2kT C_{\mu}^{2} R_q$ gives Eq. (\ref{eq14}) for $R_q$.

\end{document}